\documentclass[final]{aipproc}
\layoutstyle{6x9}

\newcommand{\Teff}{\mbox{$T_\mathrm{eff}$}}

\begin{document}

\title{Analysis of a XMM-Newton Spectrum of the Extremely Hot White Dwarf in Nova V4743\,Sgr}
      

\author{T\@. Rauch}{
  address={
  Dr.-Remeis-Sternwarte, Bamberg, University of Erlangen-N\"urnberg, Germany},
  altaddress={
  Institute for Astronomy and Astrophysics, University of T\"ubingen, 72076 T\"ubingen, Germany}
}

\author{K\@. Werner}{
  address={
  Institute for Astronomy and Astrophysics, University of T\"ubingen, 72076 T\"ubingen, Germany}
}

\author{M\@. Orio}{
  address={
  Istituto Nazionale di Astrofisica (INAF), Osservatorio Astronomico di Torino, Italy},
  altaddress={
  Department of Astronomy, University of Wisconsin, Madison, USA}
}

\begin{abstract}
Classical novae occur in close binary systems (main-sequence star + white dwarf). 
V4743\,Sgr is a very fast nova exhibiting a X-ray spectrum 
which is consistent with a view on the optically thick atmosphere of an extremely hot white dwarf (WD). 
It exhibits strong absorption features of the C\,{\sc v}, C\,{\sc vi}, N\,{\sc vi}, 
N\,{\sc vii}, and O\,{\sc vii} resonance lines 
indicating an effective temperature of about 610\,kK.

We present preliminary results of a spectral analysis by means of line-blanketed non-LTE atmosphere models 
of a XMM-Newton RGS-1 spectrum of V4743\,Sgr, taken about half a year after its outburst. 
\end{abstract}

\maketitle

\begin{figure}
 \includegraphics[width=\textwidth]{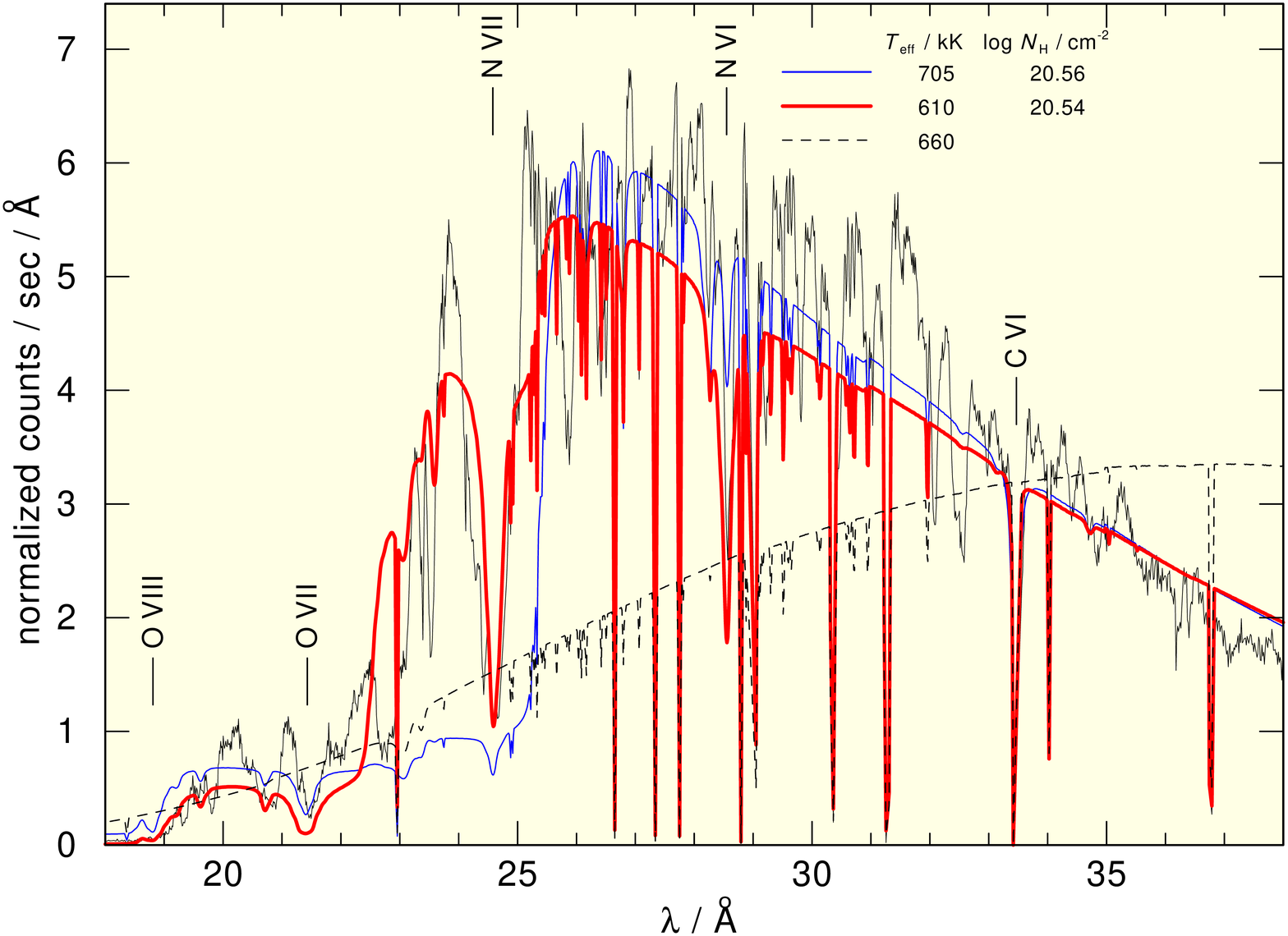} 
  \caption{
          Comparison of the XMM-Newton RGS-1 spectrum with
          NLTE model fluxes (thick: H:He:C:N:O = 18.9:75.1:0.0006:0.016:5.96, thin: solar).
          \Teff\ and $N_\mathrm{H}$ are determined with XSPEC. 
          A black body (dashed) is shown for comparison only and to identify instrumental
          features of XMM-Newton. 
          The synthetic spectra are shifted by -2400 km/sec} 
  \label{flx}
\end{figure}

\noindent
V4743\,Sgr has been discovered as a nova on Sep 20, 2002 (Haseda et al\@. 2002).
Until Mar 2003, it evolved into the brightest super-soft X-ray source in the sky.
A first CHANDRA spectrum shows a flux maximum around 30\,\AA\
and strong C\,{\sc V}, N\,{\sc VI}, and N\,{\sc VII} (Fig\@. \ref{flx})
absorption features which suggest temperatures of 1 -- 2 MK (Ness et al\@. 2003).
These lines are blue shifted (-2400\,km/sec) -- most likely we see expanding
gas which consists out of CNO-processed material.

However, in the phase after the outburst, i.e\@. when the H burning is still on-going (for years),
the surface composition of the WD is poorly known. In order to make some progress,
we decided to use our static NLTE models to investigate basic parameters like
effective temperature (\Teff) and surface composition. Although we neglect any effect of
the velocity field, our aim is at least a qualitative modeling of the 
XMM RGS-1 spectrum of V4743\,Sgr, obtained on Apr 4, 2003 with an exposure time of about 10\,h.
The EPIC-pn count rate in timing mode was 1348 cts/sec (Orio et al\@. 2003). 

For a preliminary analysis, we employed plane-parallel, static models which 
are calculated with TMAP, the T\"ubingen NLTE Model Atmosphere Package 
(Werner et al\@. 2003, Rauch \& Deetjen 2003).
The first set of models is composed of H+He+C+N+O, then Ne+Mg+S are added subsequently in
a line-formation calculation (Tab\@. \ref{stat}). However, in general TMAP can treat
all elements from hydrogen to the iron group simultaneously (Rauch 1997, 2003).

\begin{table}[t]
\caption{Statistics of the model atoms used in our calculations of the NLTE model atmospheres. 
  The notation is: NLTE = levels
  treated in NLTE, LTE = LTE levels, RBB = radiative bound-bound transitions}
\label{stat}
\smallskip
{\small
\begin{tabular}{rlrrrrlrrr}
\hline
\noalign{\smallskip}\noalign{\smallskip}
\multicolumn{1}{c}{atom} & ion & NLTE & LTE & RBB & \multicolumn{1}{c}{atom} & ion & NLTE & LTE & RBB \\
\noalign{\smallskip}
\hline
\noalign{\smallskip}
H  & {\sc i}     &   5  &  11  &  10 & Ne & {\sc vii}   &  10  &  50  &  12 \\
   & {\sc ii}    &   1  &   -  &   - &    & {\sc viii}  &   8  &  18  &  15 \\
He & {\sc i}     &   1  &  21  &   0 &    & {\sc ix}    &   5  &   6  &   3 \\
   & {\sc ii}    &  10  &  22  &  45 &    & {\sc x}     &   1  &   0  &   0 \\
   & {\sc iii}   &   1  &   -  &   - & Mg & {\sc viii}  &   1  &  19  &   0 \\
C  & {\sc v}     &  29  &  21  &  60 &    & {\sc ix}    &   3  &  23  &   1 \\
   & {\sc vi}    &  15  &  21  &  26 &    & {\sc x}     &   2  &   3  &   1 \\
   & {\sc vii}   &   1  &   0  &   - &    & {\sc xi}    &   5  &   6  &   1 \\
N  & {\sc v}     &   5  &  15  &   6 &    & {\sc xii}   &   1  &   0  &   0 \\
   & {\sc vi}    &  17  &   7  &  33 & S  & {\sc x}     &   1  &   7  &   0 \\
   & {\sc vii}   &  15  &  21  &  30 &    & {\sc xi}    &   5  &  29  &   1 \\
   & {\sc viii}  &   1  &   0  &   - &    & {\sc xii}   &  10  &  12  &  15 \\
O  & {\sc vi}    &   1  &  34  &   0 &    & {\sc xiii}  &   9  &  21  &  10 \\
   & {\sc vii}   &  19  &   7  &  21 &    & {\sc xiv}   &   9  &   1  &  15 \\
   & {\sc viii}  &  15  &  30  &  30 &    & {\sc xv}    &   1  &   0  &   0 \\
   & {\sc ix}    &   1  &   0  &   - &    &             &      &      &     \\ 
\hline
\noalign{\smallskip}\noalign{\smallskip}
   &             &      &      &     & \multicolumn{2}{r}{total} & 208 & 409 & 332 \\
\hline
\end{tabular}
}
\end{table}

Small grids of model atmospheres were calculated in order to match 
the relevant \Teff\ range (0.5 -- 1\,MK). 
208 levels of H+He+C+N+O+Ne+Mg+S are treated in NLTE with 332 individual line transitions (Tab\@. \ref{stat}).

A comparison with the spectrum of V4743\,Sgr via XSPEC yields a \Teff\ of
about 705\,kK, however, the flux at wavelengths smaller than 26\AA\ is too low.
Decreasing the C abundance by a factor of 0.01, results in a much better fit in this
region (Fig\@. \ref{flx}).

A crucial point for the understanding of processes during the nova outburst
may be the time-dependent prediction of surface abundances as well as the
spectral analysis of high-resolution X-ray spectra taken from outburst to 
the end of surface H burning. The inspection of available XMM-Newton spectra
of V4743\,Sgr, taken on a half-year time scale has shown that about weekly
spectra are highly desirable. This would also allow to investigate the velocity
reliably.  

Starrfield et al\@. (2000) calculated the yields of thermonuclear runaways of 
O/Ne/Mg WD. We adopted the average of their models' abundance ratios 
and calculated a grid of test models. Their fluxes do not reproduce the 
observation better than our simple C-reduced models. Thus, this issue will be 
a challenge for both, theory and spectral analysis in the future.

The comparison of our preliminary model atmosphere fluxes with the Apr
2003 XMM-Newton RGS spectrum of V4743\,Sgr, however,
has shown that at solar abundance ratios, we cannot fit
the overall flux distribution. Although
the fit is not perfect yet, it is obvious that the C/N abundance ratio
is much smaller (about a hundred times) than the solar value, typical
for CNO processed material. 
The fit to the observation based on our first test models
is best at a of \Teff\ about 610\,kK
and $\log N_\mathrm{H} = 20.54 \left[\mathrm{cm}^{-2}\right]$.

Analysis of X-ray data in T\"ubingen is supported by the DLR under grant 50\,OR\,0201.


\begin{thebibliography}{}

\bibitem{HEA02}  Haseda, K., West, D., Yamaoka, H., Masi, G\@. 2002, IAU Circ\@. 7975
\bibitem{KEA02}  Kato, T., Hishikura, T., West, J.D., et al\@. 2002, IAU Circ\@. 7976
\bibitem{NEA03}  Ness, J.-U., Starrfield, S., Burwitz, V., et al\@. 2003, ApJ 594, L127
\bibitem{OEA03}  Orio, M., Leibowitz, E., Rodriguez, P., et al\@. 2003, IAU Circ\@. 8131
\bibitem{R97}    Rauch, T\@. 1997, A\&A 320, 237
\bibitem{R03}    Rauch, T\@. 2003, A\&A 403, 709
\bibitem{RD03}   Rauch, T., Deetjen, J.L\@. 2003,
                 in: Workshop on Stellar Atmosphere Modeling,
                 eds\@. I\@. Hubeny, D\@. Mihalas, K\@. Werner,
                 The ASP Conference Series (San Francisco: ASP), 288, 103
\bibitem{SEA00}  Starrfield, S., Sparks, W.M., Truran J.W., Wiescher, M.C\@. 2000, ApJS 127, 485
\bibitem{WEA03}  Werner, K., Dreizler, S., Deetjen, J.L., Nagel, T., Rauch, T., Schuh, S.L\@. 2003,
                 in: Workshop on Stellar Atmosphere Modeling,
                 eds\@. I\@. Hubeny, D\@. Mihalas, K\@. Werner,
                 The ASP Conference Series (San Francisco: ASP), 288, 31

\end{thebibliography}
\end{document}